\documentclass[prb, fleqn, twocolumn]{revtex4}

\usepackage{amsmath}
\usepackage{graphicx, amssymb}
\usepackage{xcolor}
\usepackage[dvips]{epsfig}
\begin{document} 

\title{Multiple exciton generation in VO$_{2}$}
\author{S. R. Sahu$^{1}$, S. Khan$^{2}$, A. Tripathy$^{1}$, K. Dey$^{1}$, N. Bano$^{1}$, S. Raj Mohan$^{3}$, M. P. Joshi$^{3,4}$, S. Verma$^{5}$, B. T. Rao$^{5}$, V. G. Sathe$^{1}$, D. K. Shukla$^{1,*}$} 
\affiliation{$^1$ UGC-DAE Consortium for Scientific Research, Indore 452001, India\\
$^2$ Nano Science Laboratory, Materials Science Section, Raja Ramanna Centre for Advanced Technology, Indore 452013, India\\
$^3$ Photonics Nanomaterial Laboratory, laser Materials Processing Division, Raja Ramanna Centre for Advanced Technology, Indore 452013, India\\
$^4$ Homi Bhabha National Institute, Training School Complex, Anushakti Nagar, Mumbai 400094, India\\
$^5$ Laser Materials Processing Division, Raja Ramanna Centre for Advanced Technology, Indore 452013, India
}
\email{dkshukla@csr.res.in}
\date{\today}

\begin{abstract}
Multiple exciton generation (MEG) is a widely studied phenomenon in semiconductor nanocrystals and quantum dots, aimed at improving the energy conversion efficiency of solar cells. MEG is the process wherein incident photon energy is significantly larger than the band gap, and the resulting photoexcited carriers relax by generating additional electron-hole pairs, rather than decaying by heat dissipation. Here, we present an experimental demonstration of MEG in a prototype strongly correlated material, VO$_{2}$, through photocurrent spectroscopy and ultrafast transient reflectivity measurements, both of which are considered the most prominent ways for detecting MEG in working devices. The key result of this paper is the observation of MEG at room temperature (in a correlated insulating phase of VO$_{2}$), and the estimated threshold for MEG is 3E$_{g}$. We demonstrate an escalated photocurrent due to MEG in VO$_{2}$, and quantum efficiency is found to exceed 100\%. Our studies suggest that this phenomenon is a manifestation of expeditious impact ionization due to stronger electron correlations and could be exploited in a large number of strongly correlated materials. 
\end{abstract} 

\maketitle

\section{Introduction}
Optimized, large-scale manufacturing processes for the fabrication of photovoltaic devices (PVs) have promoted solar cell technologies to be on the verge of becoming the cheapest form of energy. While PVs based on materials such as silicon continue to pass economic milestones due to reduced fabrication costs and increased market penetration, the power conversion efficiency of these devices is fundamentally limited to around 33\% \cite{shockley1961detailed}. Several methods have been explored to increase the power conversion efficiency of solar cells, including the development of tandem cells, impurity bands and intermediate band devices, hot electron extraction, and carrier multiplication or multiple exciton generation (MEG) \cite{nozik2002quantum,ross1982efficiency,green2002third}. Carrier multiplication or MEG is the process in which the absorption of a single photon having energy above the threshold energy\cite{beard2011multiple, beard2013third} excites multiple electrons from the valence band to the conduction band. It has been widely studied in quantum dots, semiconductor nanocrystals, halide perovskites, etc. \cite{beard2011multiple, li2018low, timmerman2020direct, schaller2005effect, beard2008multiple, pijpers2009assessment, ten2015generating}. Some theoretical studies have predicted that materials with strong electron correlation may also exhibit MEG \cite{werner2014role,coulter2014optoelectronic, manousakis2019optimizing}. In a strongly correlated insulator (SCI), the localized electron forms an electronic system in which an effective electron-electron interaction can lead to the faster decay of initially photoexcited electron-hole pairs into multiple electron-hole pairs through a process called impact ionization (II)\cite{coulter2014optoelectronic}, resulting in an effective MEG. MEG has the potential to significantly enhance solar cell efficiency, which may lead to the realization of next generation solar cells \cite{li2018low, timmerman2020direct, schaller2005effect}. 

Manousakis $\textit{et al.}$ \cite{manousakis2010photovoltaic} theoretically studied the photovoltaic effect in narrow-gap Mott insulators and found that quantum efficiency can be significantly enhanced due to impact ionization (II) caused by photoexcited hot electron-hole pairs. Coulter $\textit{et al.}$ \cite{coulter2014optoelectronic} estimated that the II rate in a strongly correlated insulator such as VO$_2$ is approximately two orders of magnitude higher than in Si and much higher than the rate of hot electron-hole decay arising due to electron-phonon relaxation \cite{coulter2014optoelectronic}. Werner $\textit{et al.}$ \cite{werner2014role} calculated the role of II in the thermalization of photoexcited Mott insulators and emphasized that if the Mott gap is smaller than the width of the Hubbard bands, the kinetic energy of the individual carriers can be large enough to produce additional multiple carriers via a process analogous to II \cite{werner2014role}. In conventional semiconductors, the typical time-scale for II is much larger than the timescale of electron-phonon scattering, therefore the quantum efficiency (QE) value does not cross the Shockley-Queisser (SQ) efficiency limit\cite{werner2014role}. However, in SCIs, the typical time-scale for the II rate is much smaller than electron-phonon scattering \cite{shockley1961detailed}, therefore the QE may be enhanced significantly (see Fig. 1) and can overcome the SQ limit. 

VO$_2$ is one of the most widely studied strongly correlated electron materials and is famous for its near room-temperature reversible first-order insulator-to-metal transition (IMT). It transits from a low-temperature insulating monoclinic (P2$_{1}$/c) state to a high temperature metallic rutile (P4$_{2}$/mnm) state at $\sim$68 $^{\circ}$C \cite{wang2014distinct,morin1959oxides}. The IMT of VO$_2$ is one of the most intriguing phenomenon because both the structure and the electron correlation contribute to it \cite{majid2018insulator,mott1968metal,goodenough1971two,pouget1974dimerization}. A large number of experimental and theoretical investigations on VO$_{2}$ have shown the importance of strong electron correlations \cite{zylbersztejn1975metal,rice1994comment}. 

According to Goodenough, in VO$_2$ the oxygen octahedral crystal field splits the V $\textit{3d}$ degenerate orbitals into doubly degenerate $e_{g}^{\sigma}$ and triply degenerate t$_{2g}$ orbitals\cite{goodenough1971two,goodenough1973structures}. The t$_{2g}$ orbitals further split and the orbital which lies along the rutile c$_{R}$ axis ($a_{1g}$) is shifted to a lower energy relative to the other two ($e_{g}^{\pi}$) orbitals of tetragonal symmetry (of the rutile phase)\cite{goodenough1971two,goodenough1973structures}. In the insulating phase, pairing and tilting of the vanadium atoms along the rutile c$_{R}$ axis results in splitting of the $a_{1g}$ band into filled bonding and antibonding orbitals. Therefore, the insulating phase of VO$_2$ shows an optical band gap of $\sim$0.6 eV \cite{li2014ultrahigh} corresponding to the gap between the $a_{1g}$ and the $e_{g}^{\pi\ast}$ bands \cite{majid2020role,wan2017observation}. Due to its unique band structure, VO$_2$ absorbs light from the visible to the infrared (IR) range. 

\begin{figure}
\centering
\includegraphics[width=\linewidth]{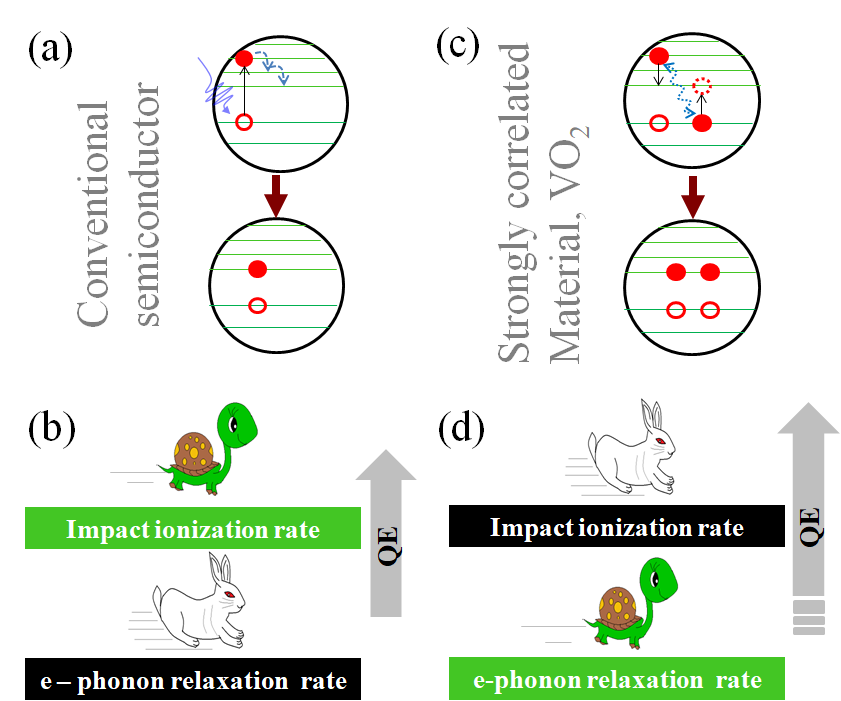}
\caption{(a) In conventional semiconductors, an incident photon with sufficient energy (h$\nu$) generates a hot carrier. The excess energy of the excited ''hot" electron beyond the energy gap is converted into waste heat by phonon emission and the electron relaxes to its band edge. (b) The typical time scale for electron-phonon relaxation is shorter than the impact ionization, so the faster electron-phonon relaxation rate in conventional semiconductors results in lower quantum efficiency. (c) In a strongly correlated material, the photo-excited electron may use its strong Coulomb interaction (shown by blue zig-zag line) with another valence electron to promote it to the conduction band, thus creating a second electron-hole pair using the energy of the same photon. (d) A faster impact ionization rate in strongly correlated materials can thus lead to higher quantum efficiency.} 
\label{F1}
\end{figure}

Major studies have used ultrafast transient absorption spectroscopy (TAS) to infer the number of electron-hole pairs produced per absorbed photon. However, due to the indirect nature of measurements and the requirement of high photon fluences, quantum yields determined from TAS have been found to be conflicting \cite{nair2011perspective, mcguire2008new, midgett2010flowing, beard2011multiple}. Furthermore, disagreements have arisen over the impact that MEG can have on solar energy conversion\cite{beard2010comparing, nair2011perspective}. Therefore, some reports have utilized the conventional photocurrent measurement method for demonstrating MEG in working devices\cite{sukhovatkin2009colloidal, semonin2011peak, kim2021escalated}.

Here, we report the observation of MEG in VO$_2$ by using a conventional photocurrent spectroscopy approach and scrutinized the underlying processes via ultrafast transient reflectivity measurements. A quantum efficiency exceeding $\sim$130\% is observed, and we are able to achieve such a high quantum efficiency under the application of small bias voltages, as suggested by earlier reports\cite{semonin2011peak, petocchi2019hund}. Photon fluence-dependent photocurrent generation showed nonlinearity, confirming a typical signature of MEG. Transient reflectivity showed Auger recombination with a typical timescale of less than 10 ps. Intensity-dependent transient reflectivity confirmed the MEG occurrence in a sample with a continuous decrement of a multiexciton lifetime with increasing intensity. Our results show that the threshold photon energy for carrier multiplication is 3E$_g$ ($\sim1.8$ eV) in VO$_2$. 

\section{Experimental details}
The VO$_2$ thin film used in this study was grown on a Si [100] substrate using the pulsed laser deposition technique. A KrF excimer laser with a wavelength of 248 nm, a repetition rate of 5Hz, and a pulse energy of 370 mJ was focused onto the V$_{2}$O$_{5}$ target with a fluence of $\approx1.1$ $J/cm^{2}$. During deposition, the ultrasonically cleaned Si substrate was maintained at a temperature of $\approx 670$ $^{\circ}$C. Deposition was performed for about 40 min in an oxygen partial pressure of $\approx8$ mTorr. A Bruker D8 x-ray diffractometer with Cu K$\alpha$ radiation was utilized to confirm the VO$_2$ phase. Temperature-dependent four-probe resistivity measurements were performed to confirm the insulator-to-metal transition in the grown VO$_2$ thin film using a home developed setup that utilized a Cryocon 22C temperature controller, a Keithley 2401 source meter, and a Keithley 2182A nanovoltmeter. Raman spectra were collected in backscattering geometry using a 10 mW Ar$^{+}$ (473 nm) laser as an excitation source coupled with a LABRAM-HR micro-Raman spectrometer equipped with a 50x objective. Absorption spectra were obtained from spectroscopic ellipsometry measurements with the help of a spectroscopic ellipsometer (model M2000, J.A. Woollam). Measurements were carried out at three different angles of incidences (50$^{\circ}$, 60$^{\circ}$, 70$^{\circ}$) in an energy range of 0.7-5 eV. The dc photocurrent measurements were performed using the Keithley 6430 signal measurement unit, and an EKSPLA NT342 Nd:YAG-based optical parametric oscillator (OPO) laser of 5 ns pulse width was used. The laser excitation energy was used in the range of 1.77-2.85 eV. 

The transient reflectivity measurement is performed using the pump-probe technique\cite{khan2014quantum, khan2015probing}. The pump-probe reflectivity measurement is done in a nondegenerate geometry. The femtosecond laser used in the pump-probe measurement is from Spectra Physics, model MAI-Tai, which is a tunable femtosecond oscillator. The pulse width is $\sim$100 fs, and the repetition rate is 80 MHz . The fundamental femtosecond laser is split into two laser pulses, the pump and the probe. The sample is excited well above 3E$_g$ at $\sim$400 nm (3.1 eV) using the second harmonic conversion of the fundamental wavelength at nearly 800 nm (1.55 eV) using one of the laser pulses (the pump pulse). The probe pulse is used for supercontinuum generation using an NKT fiber module, with which the probe wavelength peak is tuned at $\sim$617 nm (2 eV), which is near 3E$_g$ and also matches nearly with the $\textit{d}$-$\textit{d}$ transition of VO$_2$ \cite{qazilbash2008electrodynamics}. The pump and the probe beam spectrum is shown in Supplemental Fig.S2\cite{supplementary}. The detection of the time-resolved change in reflectivity $\Delta$R of the sample is measured using the standard photodiode lock-in amplifier detection technique\cite{khan2014quantum}. 

\begin{figure}
\centering
\includegraphics[width=\linewidth]{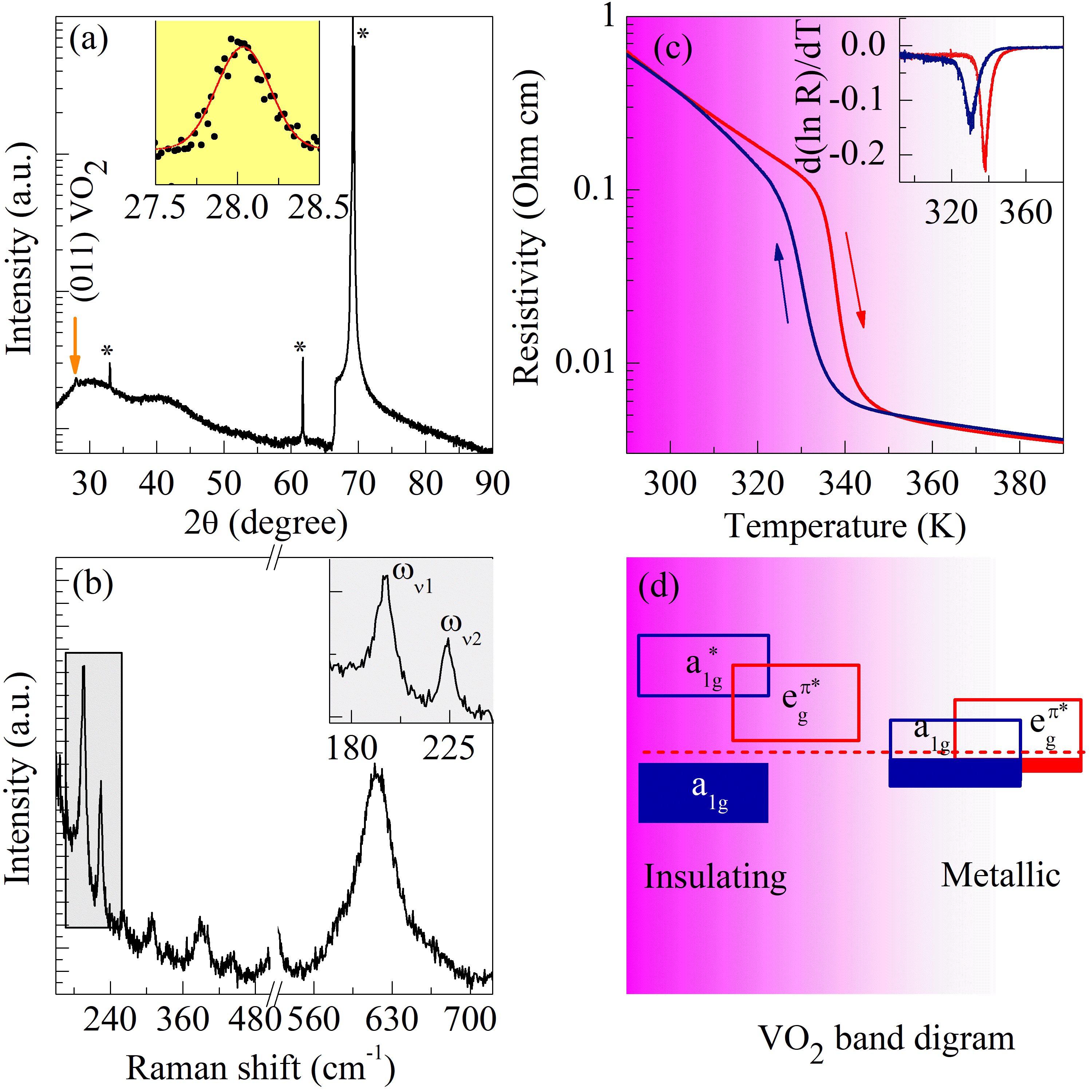}
\caption{(a) The $\theta$-2$\theta$ XRD pattern of a VO$_2$ thin film on a [100] Si substrate, with $\ast$ denoting the peaks from the Si substrate.  The inset shows the XRD peak corresponding to the (011) reflection of VO$_2$. Raman spectra of the VO$_2$ thin film is shown in (b). The inset shows the characteristic $\omega_{\upsilon1}$ and $\omega_{\upsilon2}$ modes corresponding to the V-V vibrations. (c) Temperature dependence of linear four-probe resistivity measurements in heating and cooling cycles is shown. The inset shows derivatives of the $\textit{ln R}$ vs. $\textit{T}$ plots. (d) Shows the band diagrams across the Fermi level, in insulating and metallic states of VO$_2$.}
\label{F2}
\end{figure}

\begin{figure}
\centering
\includegraphics[width=\linewidth]{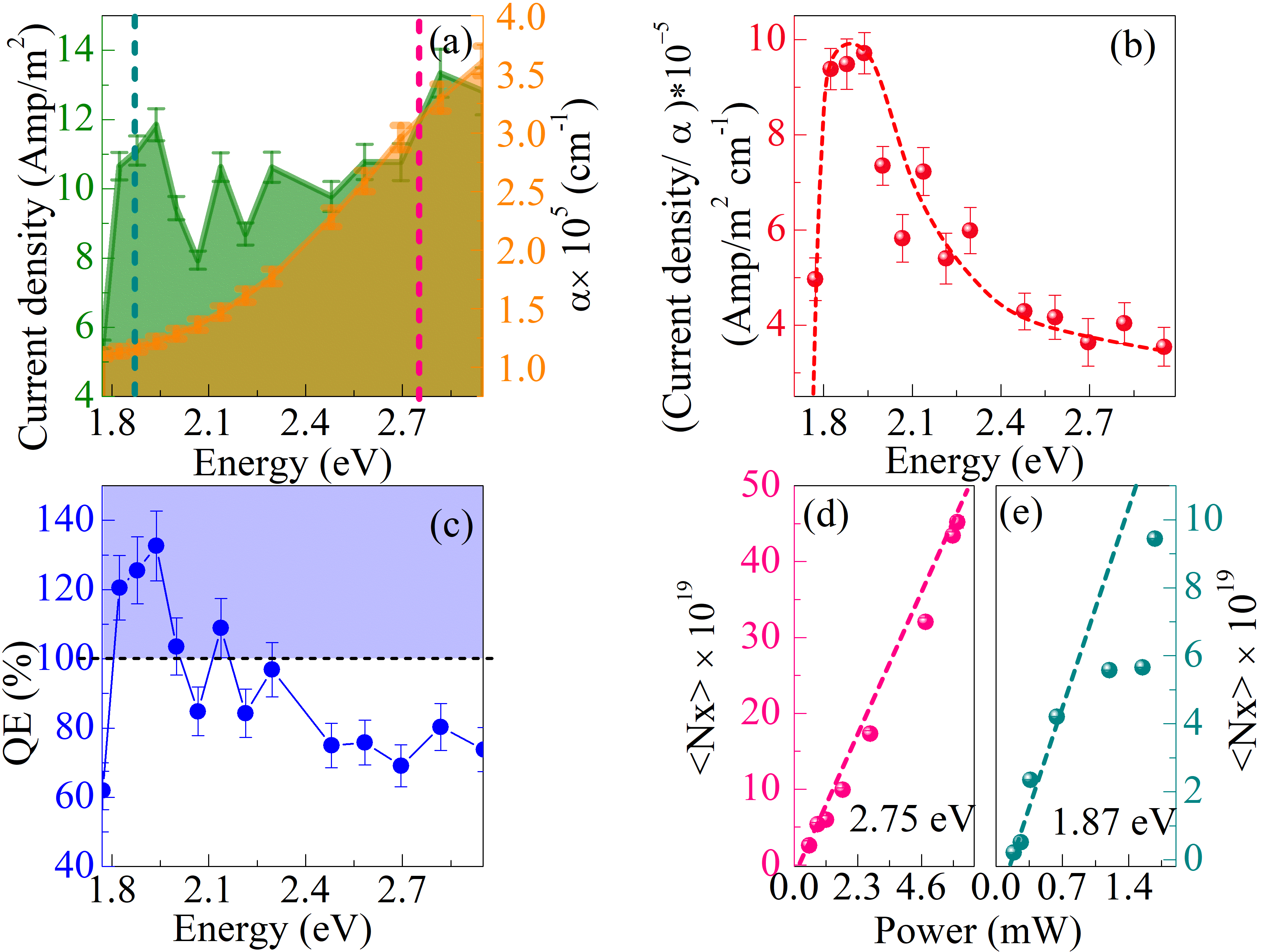}
\caption{(a) Variations of the current density and the absorption coefficient with varying photon energy. (b) To quantify the carrier multiplication, the current density is normalized by the absorption coefficient. (c) The quantum efficiency (\%) is plotted with varying photon energy, where the QE is found to exceed 100\% in the photon energy range of 1.8-2.4 eV. (d) The generated carrier flux ($<N_{x}>$) is plotted with varying intensity at (d) 2.75 eV and (e) 1.87 eV excitation energies.}
\label{F3}
\end{figure}

\section{Results and discussion}
The phase purity of the film has been confirmed by combining x-ray diffraction (XRD) and Raman spectroscopy. The XRD peak in the thin film at $\sim$28$^{\circ}$ corresponds to the (011) reflection of the VO$_2$ $\it{M1}$ phase [Fig.\ref{F2} (a)]. VO$_2$ exhibits various structural polymorphs $\cite{lee2016epitaxial}$, and XRD data of the thin film alone cannot confirm the phase; therefore, we have also performed a Raman measurement [Fig.\ref{F2} (b)]. The VO$_2$ thin film shows five $A_{g}$ and two $B_{g}$ Raman modes at $\sim$195, $\sim$223, $\sim$309, $\sim$390, and $\sim$616 cm$^{-1}$, and $\sim$261 and $\sim$441 cm$^{-1}$, respectively, which belong to the monoclinic $\it{M1}$ phase$\cite{shibuya2017polarized}$. The $A_{g}$ symmetry Raman modes $\omega$$_{\nu 1}$ and $\omega$$_{\nu 2}$ at $\sim$195 and $\sim$223 cm$^{-1}$ are assigned to the V-V vibrations (inset of Fig.\ref{F2}(b)), while the rest of the observed Raman peaks are related to V-O vibrations\cite{marini2008optical,chen2011assignment}. Figure \ref{F2}(c) shows the temperature-dependent four-probe electrical resistivity measurement of the VO$_2$ thin film. A temperature-induced phase transition from a low-temperature insulating phase to a high-temperature metallic phase is observed. The hysteresis in the resistivity data manifests the first-order nature of this transition. In order to better characterize the IMT, differential curves of Ln R vs. T are plotted (inset of Fig.\ref{F2} (c)). The IMT can be characterized by the transition temperature T$_{c}$, defined as average of the centers of the differential curves of Ln R vs T obtained during heating and cooling cycles, which is found to be $\sim$334 K. The reduced IMT temperature compared to that of the bulk VO$_2$ is attributed to the strain present in the thin film.

The current density and absorption coefficients with varying incident photon energies have been plotted in Fig.\ref{F3}(a). the photocurrent value is found to increase anomalously near $\sim$1.87 eV. The optical absorption spectra were extracted from the simultaneous fitting of the $\psi$ and $\Delta$ at all three angles (details are given in the Supplemental Material \cite{supplementary}; see also Ref. \cite{kana2011thermally} therein). Here, we are more interested in the optical band gap (E$_{g}$) of 0.6 eV for our study because this is the optical transition which is expected to be originated due to electronic correlation in VO$_2$\cite{huffman2017insulating,majid2018insulator}. In a strongly correlated insulator, the photoexcited electron (or hole) may utilize its strong Coulomb interaction with another valence electron for multiple carrier generation\cite{coulter2014optoelectronic}. To correctly estimate the generated carriers in the material, the photocurrent must be normalized by an absorption coefficient ($\alpha$). Figure \ref{F3}(b) shows the normalized photocurrent density in VO$_2$. Surprisingly, a peaklike feature in the photocurrent is observed at $\sim$1.87 eV (which is approximately three times the band gap (3E$_{g}$); E$_{g}$ $\sim$0.6 eV is the optical band gap which is expected to be originated due to electronic correlation in VO$_2$ \cite{majid2018insulator,huffman2017insulating}). Such an enhancement in photocurrent is only possible due to carrier multiplication/MEG because the number of generated carriers exceeds the number of absorbed photons. This is also supported by the quantum efficiency (QE) calculation (see Fig. \ref{F3}(c)). The QE calculation is detailed in the Supplemental Material \cite{supplementary} (see also Ref. \cite{qi2005efficient} therein). The QE exceeds 100\% for a photon energy larger than 3E$_{g}$, which gives a hint about the threshold energy requirement for the carrier multiplication in VO$_2$. Note that above 4E$_{g}$, the QE starts to decrease. A similar peak quantum efficiency exceeding 100\% in a significant photon energy range has been observed in a previous study\cite{semonin2011peak}. These photocurrent measurements have been performed at room temperature with a 0.2 V bias. The reason for using such a small amount of bias voltage is to pull out the generated charge carriers in the circuit. Earlier studies have utilized the power-dependent photocurrent measurement technique and used the slope change to quantify the carrier multiplication \cite{kim2021escalated,gao2015carrier}. We have chosen two photon energies for intensity-dependent photocurrent measurements, one $\sim$1.87 eV (where we observed the QE is exceeding by 100\%) and another $\sim$2.75 eV (where the QE is lower). Results at these two different energies have striking differences. For $\sim$2.75 eV, a linear relationship is observed between the generated carrier flux and power (Fig.\ref{F3}(d)). It shows that with increasing photon flux, the generated carrier flux is also linearly increasing. For $\sim1.87$ eV excitation energy (Fig.\ref{F3}(e)), up to a specific intensity ($\sim$0.8 mW), when the photon flux is increasing, the corresponding carrier flux is also increasing. However, when the intensity is increased further due to increased initial hot carrier density, the carrier multiplication dynamics becomes faster and the slope is suddenly changed. This is identified as the nonlinear dependence of the generated carriers on intensity (see Fig.\ref{F3}(e))\cite{villamil2020auger, kim2021escalated, gao2015carrier}.

\begin{figure}
\centering
\includegraphics[width=\linewidth]{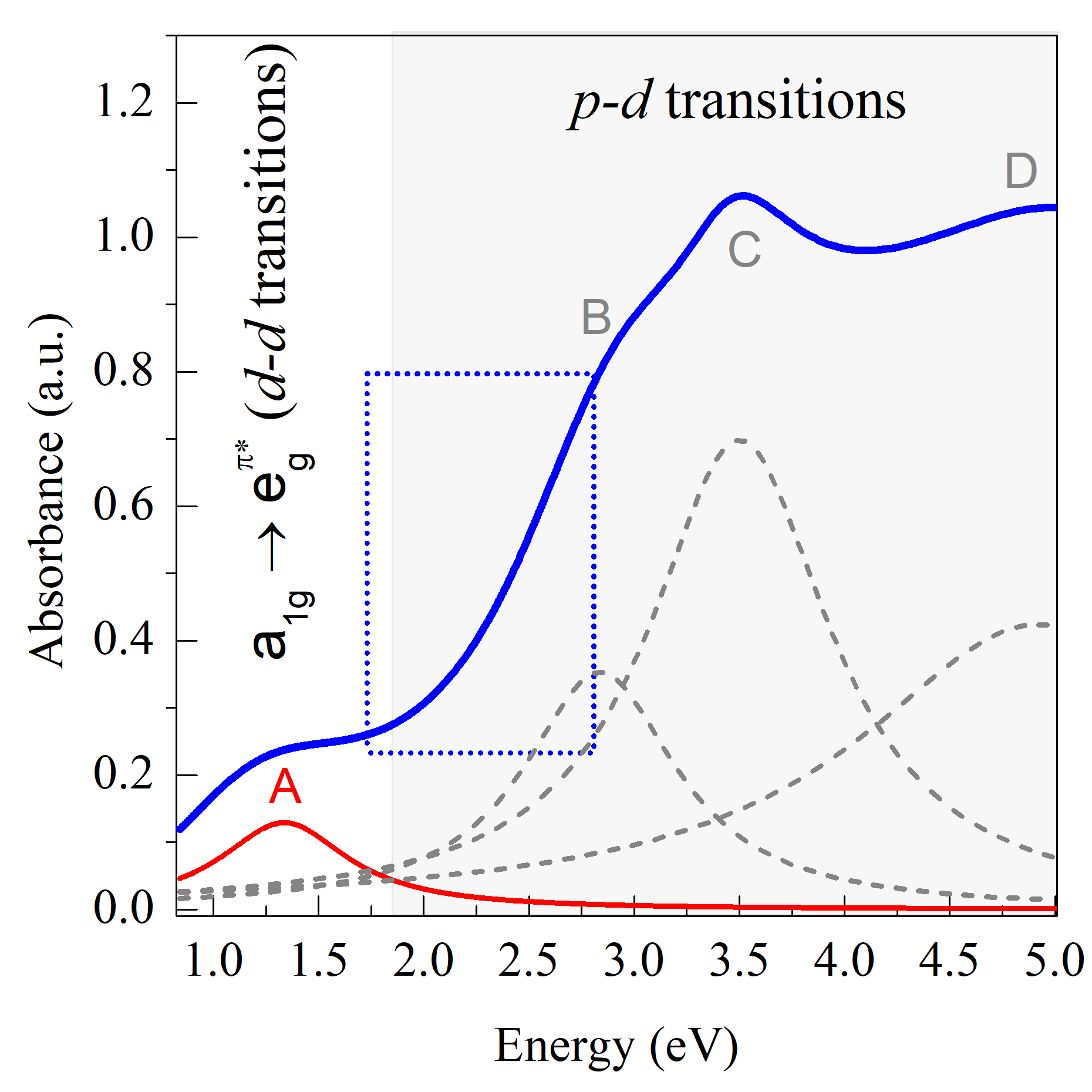}
\caption{Absorption spectrum of VO$_{2}$ plotted as a function of photon energy for the insulating phase (at room temperature). The labels in uppercase letters refer to the transitions (defined in legend) in VO$_{2}$. Dashed rectangular box depicts the range of photon energy utilized in photo-current spectroscopy measurements}
\label{F5}
\end{figure}

The behavior of the photocurrent signal and quantum efficiency with varying photon energy could be explained from the absorption processes in VO$_{2}$. The absorption spectra of VO$_{2}$ along with involved transitions (deconvoluted by utilizing an arctan background) are shown in Fig. \ref{F5}. The hump feature in Fig. \ref{F5} labeled as ''A” at $\sim$1.3 eV is due to an optical transition from the filled lower a$_{1g}$  band to the empty $e_{g}^{\pi\ast}$ bands ($\textit{d}$-$\textit{d}$ transition) across an optical band gap (E$_{g}$) of $\sim$0.6 eV. All other features ''B”, ''C” and ''D” (at $\sim$2.8 eV, $\sim$3.5 eV, and $\sim$4.9 eV)  are attributed to $\textit{p}$-$\textit{d}$ transitions \cite{caruthers1973energy,gavini1972optical,verleur1968optical,dai2019optical,schneider2020optical}. We have observed the peaklike feature in the normalized current density at $\sim$1.87 eV and above this photon energy the photocurrent signal is decreasing. At the lower photon energies ($\le$1.87 eV) there is only one optical transition which occurs, i.e., the a$_{1g}$ to $e_{g}^{\pi\ast}$ and $\textit{n}$E$_{g}$ (n \textgreater 2) condition required for MEG is satisfied. At higher photon energies (\textgreater1.87 eV) B/C transitions involving $\textit{p}$-$\textit{d}$ transitions start to occur \cite{gavini1972optical,schneider2020optical} due to which the hot carrier generation and impact ionization will stop. It is also to be noted that the theoretical calculations have suggested that highly correlated bands originate from vanadium $\textit{d}$ orbitals, while the conventional bands are made up of oxygen $\textit{p}$ orbitals \cite{coulter2014optoelectronic}. The generated multiexciton due to impact ionization is associated with the subbands of $\textit{d}$ electrons (correlated electrons), while $\textit{p}$ bands show similar character as conventional semiconductor. Moreover, Coulter $et$ $al.$, theoretically demonstrated that the highly correlated electrons provide a high impact ionization rate, which is the basis of multiple exciton generation in VO$_{2}$ \cite{coulter2014optoelectronic}. 

\begin{figure}
\centering
\includegraphics[width=\linewidth]{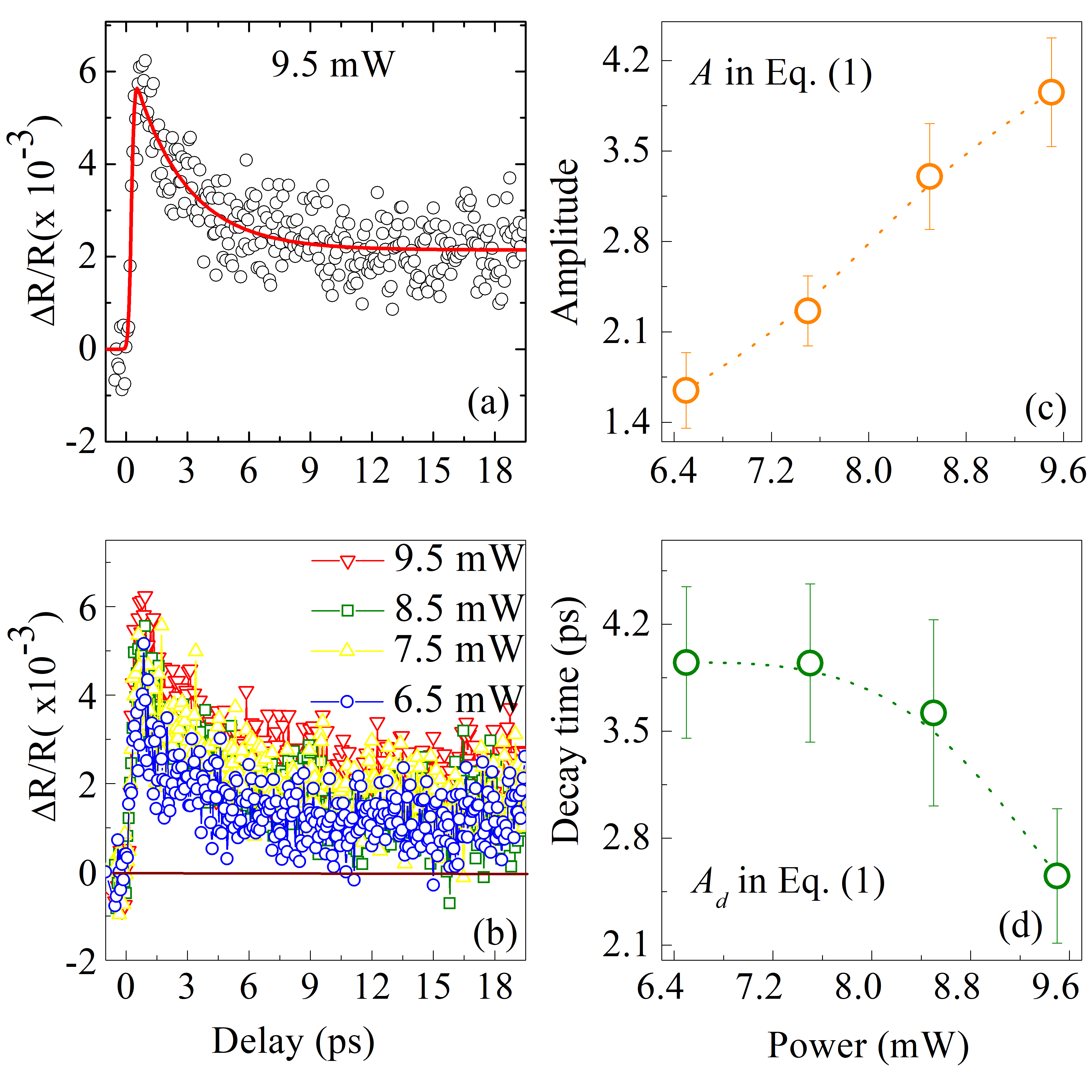}
\caption{(a) Fitted differential transient reflectivity curve (red curve) at a fluence of 9.5 mW with raw data (purple circles). (b) Power dependence of differential transient reflectivity curves. (c)Amplitude and (d) decay time obtained from fitting the transient reflectivity data (Eq. (1)) as a function of fluence.}
\label{F4}
\end{figure}

 In SCIs the impact ionization process occurs due to the presence of a strong electron-electron interaction. Therefore initially a photogenerated exciton induces the generation of a multiexciton. Finally, the generated multiexciton via impact ionization can relax through the nonradiative Auger recombination (AR) process, which is the most dominant pathway for multiexciton recombination. To study the recombination dynamics in an ultrafast time scale in VO$_2$, we tracked the AR dynamics through the transient reflectivity measurement, which serves as a signature for MEG. To investigate the AR process in VO$_2$, we used a pump beam of 3.1 eV excitation energy, which is sufficiently larger than the band gap of VO$_2$. As discussed earlier, we obtained a maximum efficiency at $\sim$1.87 eV in a photocurrent measurement, so to support our photocurrent spectroscopy data we explored the AR dynamics near this energy range. We utilized white light as a probe beam, with a maximum intensity at $\sim$2 eV (pump and probe beam spectra can be seen in Figs. S2(a) and S2(b), respectively) which is nearer to $\sim$1.87 eV. Due to the experimental limitation, we could only probe near 2 eV, and not beyond that. Interestingly, with the help of this pump and probe beam, we obtained clear signatures of the decay of the photoexcited carriers in VO$_2$, which is discussed below. AR is the inverse process of impact ionization and has been characterized by a fast intrinsic AR rate with a corresponding effective recombination time of less than 10 ps\cite{villamil2020auger}. Depending on the nature of the photoexcited species or initial hot carrier density, the AR process follows second or third order kinetics\cite{pelant2012luminescence, villamil2020auger}. The transient reflectivity data have been fitted with a combination of an error function and exponential decaying function,
\begin{equation}
\frac{\Delta R}{R} = {0.5}{\ast (erf({\frac{(t-t_o)}{S_r}})+1)}{{\ast (A}{\ast exp{(\frac{-(t-t_o)}{A_d})}+B})}\;
\end{equation}

with 95\% confidence bounds. An error function has been used to take into account the rise time of the carrier excitation (S$_{r}$) and the exponential decay function (A$_{d}$) gives the details of the initial fast decay time of the carriers. Our transient differential reflectivity ($\Delta$R) fitting results also clearly indicate that Auger recombination has occurred in less than 10 ps (Figs. \ref{F4}(a) and \ref{F4}(b)) at each fluence. The valuable parameters such as rise time (S$_{r}$), amplitude ($\textit{A}$), decay time (A$_{d}$)), and background ($\textit{B}$) are extracted from the fitting using the above discussed equation (see the representative fitting for 9.5 mW in Fig. \ref{F4}(a)). In Fig. \ref{F4}(c), increasing amplitude with increasing fluence represents the enhancement of the photoexcited carrier density with fluence. The decrement of the decay time (A$_{d}$) with increasing fluence (see Fig. \ref{F4}(d)) might be a fingerprint of the AR which results in  MEG \cite{villamil2020auger,werner2014role}. Moreover, the experimental dynamics become faster with an increase in the initial carrier density, which is in agreement with the enhancement of the effective linear AR rates with the excitation fluence. This expresses a nonlinear dependence of the kinetics related to the multiexciton dynamics depending on the exciton density. An almost similar background has been observed at each power for the transient reflectivity data (see Table S1) which can be explained with the help of a similar timescale of the single excitonic decay\cite{villamil2020auger,holleman2016evidence}. We have fitted with the one exponential decay function because the measurement was performed for a small timescale, up to 120 ps. If one measures for longer delays, then information related to the relaxed carrier features can also be obtained.	

\section{Conclusions}
 In conclusion, photoinduced MEG has been investigated in the correlated electron system VO$_2$ by means of a direct method of dc photocurrent spectroscopy and an indirect method of ultrafast transient reflectivity. Photocurrent spectroscopy exhibits an enhanced photocurrent and the resultant QE is found to increase up to as high as $\sim$130\%. A nonlinear dependence of the generated carrier flux with increasing photon flux clearly indicates MEG in VO$_2$. Photon fluence-dependent ultrafast transient reflectivity measurements show a decrement of Auger recombination time from 3.95 to 2.4 ps with increasing photon fluence, which is the most prominent pathway for multiple exciton recombination as the exciton density increases. This is accompanied by an amplitude enhancement from 1.6 to 3.96 by increasing the fluence from 6.5 to 9.5 mW. These results are manifestation that Auger recombination is occuring within the VO$_2$ thin film and complements to the MEG findings in photocurrent spectroscopy. Our photocurrent results confirm that the MEG threshold for VO$_2$ is 3E$_{g}$. In the case of VO$_2$, the MEG threshold may vary with temperature because the band gap and correlation strength decrease with increasing temperature \cite{okazaki2002temperature, wang2015distinguishing}. This study may set off an era of strongly correlated material-based high-performance optoelectronic devices, such as solar cells and photo-detectors, in the near future. 
 
 \section{Acknowledgements}
 D.K.S. acknowledges support from SERB New Delhi, India in the form of early career research award (ECR/2017/000712).
\bibliography{VO2SiphotoconductivityF}

\begin{thebibliography}{58}
\expandafter\ifx\csname natexlab\endcsname\relax\def\natexlab#1{#1}\fi
\expandafter\ifx\csname bibnamefont\endcsname\relax
  \def\bibnamefont#1{#1}\fi
\expandafter\ifx\csname bibfnamefont\endcsname\relax
  \def\bibfnamefont#1{#1}\fi
\expandafter\ifx\csname citenamefont\endcsname\relax
  \def\citenamefont#1{#1}\fi
\expandafter\ifx\csname url\endcsname\relax
  \def\url#1{\texttt{#1}}\fi
\expandafter\ifx\csname urlprefix\endcsname\relax\def\urlprefix{URL }\fi
\providecommand{\bibinfo}[2]{#2}
\providecommand{\eprint}[2][]{\url{#2}}

\bibitem[{\citenamefont{Shockley and Queisser}(1961)}]{shockley1961detailed}
\bibinfo{author}{\bibfnamefont{W.}~\bibnamefont{Shockley}} \bibnamefont{and} \bibinfo{author}{\bibfnamefont{H.~J.} \bibnamefont{Queisser}}, \bibinfo{journal}{Journal of applied physics} \textbf{\bibinfo{volume}{32}}, \bibinfo{pages}{510} (\bibinfo{year}{1961}).

\bibitem[{\citenamefont{Nozik}(2002)}]{nozik2002quantum}
\bibinfo{author}{\bibfnamefont{A.~J.} \bibnamefont{Nozik}}, \bibinfo{journal}{Physica E: Low-dimensional Systems and Nanostructures} \textbf{\bibinfo{volume}{14}}, \bibinfo{pages}{115} (\bibinfo{year}{2002}).

\bibitem[{\citenamefont{Ross and Nozik}(1982)}]{ross1982efficiency}
\bibinfo{author}{\bibfnamefont{R.~T.} \bibnamefont{Ross}} \bibnamefont{and} \bibinfo{author}{\bibfnamefont{A.~J.} \bibnamefont{Nozik}}, \bibinfo{journal}{Journal of Applied Physics} \textbf{\bibinfo{volume}{53}}, \bibinfo{pages}{3813} (\bibinfo{year}{1982}).

\bibitem[{\citenamefont{Green}(2002)}]{green2002third}
\bibinfo{author}{\bibfnamefont{M.~A.} \bibnamefont{Green}}, \bibinfo{journal}{Physica E: Low-dimensional Systems and Nanostructures} \textbf{\bibinfo{volume}{14}}, \bibinfo{pages}{65} (\bibinfo{year}{2002}).

\bibitem[{\citenamefont{Beard}(2011)}]{beard2011multiple}
\bibinfo{author}{\bibfnamefont{M.~C.} \bibnamefont{Beard}}, \bibinfo{journal}{The Journal of Physical Chemistry Letters} \textbf{\bibinfo{volume}{2}}, \bibinfo{pages}{1282} (\bibinfo{year}{2011}).

\bibitem[{\citenamefont{Beard et~al.}(2013)\citenamefont{Beard, Luther, Semonin, and Nozik}}]{beard2013third}
\bibinfo{author}{\bibfnamefont{M.~C.} \bibnamefont{Beard}}, \bibinfo{author}{\bibfnamefont{J.~M.} \bibnamefont{Luther}}, \bibinfo{author}{\bibfnamefont{O.~E.} \bibnamefont{Semonin}}, \bibnamefont{and} \bibinfo{author}{\bibfnamefont{A.~J.} \bibnamefont{Nozik}}, \bibinfo{journal}{Accounts of chemical research} \textbf{\bibinfo{volume}{46}}, \bibinfo{pages}{1252} (\bibinfo{year}{2013}).

\bibitem[{\citenamefont{Li et~al.}(2018)\citenamefont{Li, Begum, Fu, Xu, Koh, Veldhuis, Gr{\"a}tzel, Mathews, Mhaisalkar, and Sum}}]{li2018low}
\bibinfo{author}{\bibfnamefont{M.}~\bibnamefont{Li}}, \bibinfo{author}{\bibfnamefont{R.}~\bibnamefont{Begum}}, \bibinfo{author}{\bibfnamefont{J.}~\bibnamefont{Fu}}, \bibinfo{author}{\bibfnamefont{Q.}~\bibnamefont{Xu}}, \bibinfo{author}{\bibfnamefont{T.~M.} \bibnamefont{Koh}}, \bibinfo{author}{\bibfnamefont{S.~A.} \bibnamefont{Veldhuis}}, \bibinfo{author}{\bibfnamefont{M.}~\bibnamefont{Gr{\"a}tzel}}, \bibinfo{author}{\bibfnamefont{N.}~\bibnamefont{Mathews}}, \bibinfo{author}{\bibfnamefont{S.}~\bibnamefont{Mhaisalkar}}, \bibnamefont{and} \bibinfo{author}{\bibfnamefont{T.~C.} \bibnamefont{Sum}}, \bibinfo{journal}{Nature communications} \textbf{\bibinfo{volume}{9}}, \bibinfo{pages}{1} (\bibinfo{year}{2018}).

\bibitem[{\citenamefont{Timmerman et~al.}(2020)\citenamefont{Timmerman, Matsubara, Gomez, Ashida, Gregorkiewicz, and Fujiwara}}]{timmerman2020direct}
\bibinfo{author}{\bibfnamefont{D.}~\bibnamefont{Timmerman}}, \bibinfo{author}{\bibfnamefont{E.}~\bibnamefont{Matsubara}}, \bibinfo{author}{\bibfnamefont{L.}~\bibnamefont{Gomez}}, \bibinfo{author}{\bibfnamefont{M.}~\bibnamefont{Ashida}}, \bibinfo{author}{\bibfnamefont{T.}~\bibnamefont{Gregorkiewicz}}, \bibnamefont{and} \bibinfo{author}{\bibfnamefont{Y.}~\bibnamefont{Fujiwara}}, \bibinfo{journal}{ACS omega} \textbf{\bibinfo{volume}{5}}, \bibinfo{pages}{21506} (\bibinfo{year}{2020}).

\bibitem[{\citenamefont{Schaller et~al.}(2005)\citenamefont{Schaller, Petruska, and Klimov}}]{schaller2005effect}
\bibinfo{author}{\bibfnamefont{R.~D.} \bibnamefont{Schaller}}, \bibinfo{author}{\bibfnamefont{M.~A.} \bibnamefont{Petruska}}, \bibnamefont{and} \bibinfo{author}{\bibfnamefont{V.~I.} \bibnamefont{Klimov}}, \bibinfo{journal}{Applied Physics Letters} \textbf{\bibinfo{volume}{87}}, \bibinfo{pages}{253102} (\bibinfo{year}{2005}).

\bibitem[{\citenamefont{Beard and Ellingson}(2008)}]{beard2008multiple}
\bibinfo{author}{\bibfnamefont{M.~C.} \bibnamefont{Beard}} \bibnamefont{and} \bibinfo{author}{\bibfnamefont{R.~J.} \bibnamefont{Ellingson}}, \bibinfo{journal}{Laser \& Photonics Reviews} \textbf{\bibinfo{volume}{2}}, \bibinfo{pages}{377} (\bibinfo{year}{2008}).

\bibitem[{\citenamefont{Pijpers et~al.}(2009)\citenamefont{Pijpers, Ulbricht, Tielrooij, Osherov, Golan, Delerue, Allan, and Bonn}}]{pijpers2009assessment}
\bibinfo{author}{\bibfnamefont{J.}~\bibnamefont{Pijpers}}, \bibinfo{author}{\bibfnamefont{R.}~\bibnamefont{Ulbricht}}, \bibinfo{author}{\bibfnamefont{K.}~\bibnamefont{Tielrooij}}, \bibinfo{author}{\bibfnamefont{A.}~\bibnamefont{Osherov}}, \bibinfo{author}{\bibfnamefont{Y.}~\bibnamefont{Golan}}, \bibinfo{author}{\bibfnamefont{C.}~\bibnamefont{Delerue}}, \bibinfo{author}{\bibfnamefont{G.}~\bibnamefont{Allan}}, \bibnamefont{and} \bibinfo{author}{\bibfnamefont{M.}~\bibnamefont{Bonn}}, \bibinfo{journal}{Nature Physics} \textbf{\bibinfo{volume}{5}}, \bibinfo{pages}{811} (\bibinfo{year}{2009}).

\bibitem[{\citenamefont{Ten~Cate et~al.}(2015)\citenamefont{Ten~Cate, Sandeep, Liu, Law, Kinge, Houtepen, Schins, and Siebbeles}}]{ten2015generating}
\bibinfo{author}{\bibfnamefont{S.}~\bibnamefont{Ten~Cate}}, \bibinfo{author}{\bibfnamefont{C.~S.} \bibnamefont{Sandeep}}, \bibinfo{author}{\bibfnamefont{Y.}~\bibnamefont{Liu}}, \bibinfo{author}{\bibfnamefont{M.}~\bibnamefont{Law}}, \bibinfo{author}{\bibfnamefont{S.}~\bibnamefont{Kinge}}, \bibinfo{author}{\bibfnamefont{A.~J.} \bibnamefont{Houtepen}}, \bibinfo{author}{\bibfnamefont{J.~M.} \bibnamefont{Schins}}, \bibnamefont{and} \bibinfo{author}{\bibfnamefont{L.~D.} \bibnamefont{Siebbeles}}, \bibinfo{journal}{Accounts of chemical research} \textbf{\bibinfo{volume}{48}}, \bibinfo{pages}{174} (\bibinfo{year}{2015}).

\bibitem[{\citenamefont{Werner et~al.}(2014)\citenamefont{Werner, Held, and Eckstein}}]{werner2014role}
\bibinfo{author}{\bibfnamefont{P.}~\bibnamefont{Werner}}, \bibinfo{author}{\bibfnamefont{K.}~\bibnamefont{Held}}, \bibnamefont{and} \bibinfo{author}{\bibfnamefont{M.}~\bibnamefont{Eckstein}}, \bibinfo{journal}{Physical Review B} \textbf{\bibinfo{volume}{90}}, \bibinfo{pages}{235102} (\bibinfo{year}{2014}).

\bibitem[{\citenamefont{Coulter et~al.}(2014)\citenamefont{Coulter, Manousakis, and Gali}}]{coulter2014optoelectronic}
\bibinfo{author}{\bibfnamefont{J.~E.} \bibnamefont{Coulter}}, \bibinfo{author}{\bibfnamefont{E.}~\bibnamefont{Manousakis}}, \bibnamefont{and} \bibinfo{author}{\bibfnamefont{A.}~\bibnamefont{Gali}}, \bibinfo{journal}{Physical Review B} \textbf{\bibinfo{volume}{90}}, \bibinfo{pages}{165142} (\bibinfo{year}{2014}).

\bibitem[{\citenamefont{Manousakis}(2019)}]{manousakis2019optimizing}
\bibinfo{author}{\bibfnamefont{E.}~\bibnamefont{Manousakis}}, \bibinfo{journal}{Scientific reports} \textbf{\bibinfo{volume}{9}}, \bibinfo{pages}{1} (\bibinfo{year}{2019}).

\bibitem[{\citenamefont{Manousakis}(2010)}]{manousakis2010photovoltaic}
\bibinfo{author}{\bibfnamefont{E.}~\bibnamefont{Manousakis}}, \bibinfo{journal}{Physical Review B} \textbf{\bibinfo{volume}{82}}, \bibinfo{pages}{125109} (\bibinfo{year}{2010}).

\bibitem[{\citenamefont{Wang et~al.}(2014)\citenamefont{Wang, Novikova, Klopf, Madaras, Williams, Madaras, Lu, Wolf, and Lukaszew}}]{wang2014distinct}
\bibinfo{author}{\bibfnamefont{L.}~\bibnamefont{Wang}}, \bibinfo{author}{\bibfnamefont{I.}~\bibnamefont{Novikova}}, \bibinfo{author}{\bibfnamefont{J.~M.} \bibnamefont{Klopf}}, \bibinfo{author}{\bibfnamefont{S.}~\bibnamefont{Madaras}}, \bibinfo{author}{\bibfnamefont{G.~P.} \bibnamefont{Williams}}, \bibinfo{author}{\bibfnamefont{E.}~\bibnamefont{Madaras}}, \bibinfo{author}{\bibfnamefont{J.}~\bibnamefont{Lu}}, \bibinfo{author}{\bibfnamefont{S.~A.} \bibnamefont{Wolf}}, \bibnamefont{and} \bibinfo{author}{\bibfnamefont{R.~A.} \bibnamefont{Lukaszew}}, \bibinfo{journal}{Advanced Optical Materials} \textbf{\bibinfo{volume}{2}}, \bibinfo{pages}{30} (\bibinfo{year}{2014}).

\bibitem[{\citenamefont{Morin}(1959)}]{morin1959oxides}
\bibinfo{author}{\bibfnamefont{F.}~\bibnamefont{Morin}}, \bibinfo{journal}{Physical review letters} \textbf{\bibinfo{volume}{3}}, \bibinfo{pages}{34} (\bibinfo{year}{1959}).

\bibitem[{\citenamefont{Majid et~al.}(2018)\citenamefont{Majid, Shukla, Rahman, Khan, Gautam, Ahad, Francoual, Choudhary, Sathe, and Strempfer}}]{majid2018insulator}
\bibinfo{author}{\bibfnamefont{S.}~\bibnamefont{Majid}}, \bibinfo{author}{\bibfnamefont{D.}~\bibnamefont{Shukla}}, \bibinfo{author}{\bibfnamefont{F.}~\bibnamefont{Rahman}}, \bibinfo{author}{\bibfnamefont{S.}~\bibnamefont{Khan}}, \bibinfo{author}{\bibfnamefont{K.}~\bibnamefont{Gautam}}, \bibinfo{author}{\bibfnamefont{A.}~\bibnamefont{Ahad}}, \bibinfo{author}{\bibfnamefont{S.}~\bibnamefont{Francoual}}, \bibinfo{author}{\bibfnamefont{R.}~\bibnamefont{Choudhary}}, \bibinfo{author}{\bibfnamefont{V.}~\bibnamefont{Sathe}}, \bibnamefont{and} \bibinfo{author}{\bibfnamefont{J.}~\bibnamefont{Strempfer}}, \bibinfo{journal}{Physical Review B} \textbf{\bibinfo{volume}{98}}, \bibinfo{pages}{075152} (\bibinfo{year}{2018}).

\bibitem[{\citenamefont{Mott}(1968)}]{mott1968metal}
\bibinfo{author}{\bibfnamefont{N.~F.} \bibnamefont{Mott}}, \bibinfo{journal}{Reviews of Modern Physics} \textbf{\bibinfo{volume}{40}}, \bibinfo{pages}{677} (\bibinfo{year}{1968}).

\bibitem[{\citenamefont{Goodenough}(1971)}]{goodenough1971two}
\bibinfo{author}{\bibfnamefont{J.~B.} \bibnamefont{Goodenough}}, \bibinfo{journal}{Journal of Solid State Chemistry} \textbf{\bibinfo{volume}{3}}, \bibinfo{pages}{490} (\bibinfo{year}{1971}).

\bibitem[{\citenamefont{Pouget et~al.}(1974)\citenamefont{Pouget, Launois, Rice, Dernier, Gossard, Villeneuve, and Hagenmuller}}]{pouget1974dimerization}
\bibinfo{author}{\bibfnamefont{J.}~\bibnamefont{Pouget}}, \bibinfo{author}{\bibfnamefont{H.}~\bibnamefont{Launois}}, \bibinfo{author}{\bibfnamefont{T.}~\bibnamefont{Rice}}, \bibinfo{author}{\bibfnamefont{P.}~\bibnamefont{Dernier}}, \bibinfo{author}{\bibfnamefont{A.}~\bibnamefont{Gossard}}, \bibinfo{author}{\bibfnamefont{G.}~\bibnamefont{Villeneuve}}, \bibnamefont{and} \bibinfo{author}{\bibfnamefont{P.}~\bibnamefont{Hagenmuller}}, \bibinfo{journal}{Physical Review B} \textbf{\bibinfo{volume}{10}}, \bibinfo{pages}{1801} (\bibinfo{year}{1974}).

\bibitem[{\citenamefont{Zylbersztejn and Mott}(1975)}]{zylbersztejn1975metal}
\bibinfo{author}{\bibfnamefont{A.}~\bibnamefont{Zylbersztejn}} \bibnamefont{and} \bibinfo{author}{\bibfnamefont{N.~F.} \bibnamefont{Mott}}, \bibinfo{journal}{Physical Review B} \textbf{\bibinfo{volume}{11}}, \bibinfo{pages}{4383} (\bibinfo{year}{1975}).

\bibitem[{\citenamefont{Rice et~al.}(1994)\citenamefont{Rice, Launois, and Pouget}}]{rice1994comment}
\bibinfo{author}{\bibfnamefont{T.}~\bibnamefont{Rice}}, \bibinfo{author}{\bibfnamefont{H.}~\bibnamefont{Launois}}, \bibnamefont{and} \bibinfo{author}{\bibfnamefont{J.}~\bibnamefont{Pouget}}, \bibinfo{journal}{Physical review letters} \textbf{\bibinfo{volume}{73}}, \bibinfo{pages}{3042} (\bibinfo{year}{1994}).

\bibitem[{\citenamefont{Goodenough and Hong}(1973)}]{goodenough1973structures}
\bibinfo{author}{\bibfnamefont{J.}~\bibnamefont{Goodenough}} \bibnamefont{and} \bibinfo{author}{\bibfnamefont{H.~Y.} \bibnamefont{Hong}}, \bibinfo{journal}{Physical Review B} \textbf{\bibinfo{volume}{8}}, \bibinfo{pages}{1323} (\bibinfo{year}{1973}).

\bibitem[{\citenamefont{Li et~al.}(2014)\citenamefont{Li, Hu, Peng, Wu, Yang, Feng, Gao, Yang, and Xie}}]{li2014ultrahigh}
\bibinfo{author}{\bibfnamefont{Z.}~\bibnamefont{Li}}, \bibinfo{author}{\bibfnamefont{Z.}~\bibnamefont{Hu}}, \bibinfo{author}{\bibfnamefont{J.}~\bibnamefont{Peng}}, \bibinfo{author}{\bibfnamefont{C.}~\bibnamefont{Wu}}, \bibinfo{author}{\bibfnamefont{Y.}~\bibnamefont{Yang}}, \bibinfo{author}{\bibfnamefont{F.}~\bibnamefont{Feng}}, \bibinfo{author}{\bibfnamefont{P.}~\bibnamefont{Gao}}, \bibinfo{author}{\bibfnamefont{J.}~\bibnamefont{Yang}}, \bibnamefont{and} \bibinfo{author}{\bibfnamefont{Y.}~\bibnamefont{Xie}}, \bibinfo{journal}{Advanced Functional Materials} \textbf{\bibinfo{volume}{24}}, \bibinfo{pages}{1821} (\bibinfo{year}{2014}).

\bibitem[{\citenamefont{Majid et~al.}(2020)\citenamefont{Majid, Sahu, Ahad, Dey, Gautam, Rahman, Behera, Deshpande, Sathe, and Shukla}}]{majid2020role}
\bibinfo{author}{\bibfnamefont{S.}~\bibnamefont{Majid}}, \bibinfo{author}{\bibfnamefont{S.}~\bibnamefont{Sahu}}, \bibinfo{author}{\bibfnamefont{A.}~\bibnamefont{Ahad}}, \bibinfo{author}{\bibfnamefont{K.}~\bibnamefont{Dey}}, \bibinfo{author}{\bibfnamefont{K.}~\bibnamefont{Gautam}}, \bibinfo{author}{\bibfnamefont{F.}~\bibnamefont{Rahman}}, \bibinfo{author}{\bibfnamefont{P.}~\bibnamefont{Behera}}, \bibinfo{author}{\bibfnamefont{U.}~\bibnamefont{Deshpande}}, \bibinfo{author}{\bibfnamefont{V.}~\bibnamefont{Sathe}}, \bibnamefont{and} \bibinfo{author}{\bibfnamefont{D.}~\bibnamefont{Shukla}}, \bibinfo{journal}{Physical Review B} \textbf{\bibinfo{volume}{101}}, \bibinfo{pages}{014108} (\bibinfo{year}{2020}).

\bibitem[{\citenamefont{Wan et~al.}(2017)\citenamefont{Wan, Xiong, Li, Liu, Wang, Ching, and Zhao}}]{wan2017observation}
\bibinfo{author}{\bibfnamefont{M.}~\bibnamefont{Wan}}, \bibinfo{author}{\bibfnamefont{M.}~\bibnamefont{Xiong}}, \bibinfo{author}{\bibfnamefont{N.}~\bibnamefont{Li}}, \bibinfo{author}{\bibfnamefont{B.}~\bibnamefont{Liu}}, \bibinfo{author}{\bibfnamefont{S.}~\bibnamefont{Wang}}, \bibinfo{author}{\bibfnamefont{W.-Y.} \bibnamefont{Ching}}, \bibnamefont{and} \bibinfo{author}{\bibfnamefont{X.}~\bibnamefont{Zhao}}, \bibinfo{journal}{Applied Surface Science} \textbf{\bibinfo{volume}{410}}, \bibinfo{pages}{363} (\bibinfo{year}{2017}).

\bibitem[{\citenamefont{Nair et~al.}(2011)\citenamefont{Nair, Chang, Geyer, and Bawendi}}]{nair2011perspective}
\bibinfo{author}{\bibfnamefont{G.}~\bibnamefont{Nair}}, \bibinfo{author}{\bibfnamefont{L.-Y.} \bibnamefont{Chang}}, \bibinfo{author}{\bibfnamefont{S.~M.} \bibnamefont{Geyer}}, \bibnamefont{and} \bibinfo{author}{\bibfnamefont{M.~G.} \bibnamefont{Bawendi}}, \bibinfo{journal}{Nano letters} \textbf{\bibinfo{volume}{11}}, \bibinfo{pages}{2145} (\bibinfo{year}{2011}).

\bibitem[{\citenamefont{McGuire et~al.}(2008)\citenamefont{McGuire, Joo, Pietryga, Schaller, and Klimov}}]{mcguire2008new}
\bibinfo{author}{\bibfnamefont{J.~A.} \bibnamefont{McGuire}}, \bibinfo{author}{\bibfnamefont{J.}~\bibnamefont{Joo}}, \bibinfo{author}{\bibfnamefont{J.~M.} \bibnamefont{Pietryga}}, \bibinfo{author}{\bibfnamefont{R.~D.} \bibnamefont{Schaller}}, \bibnamefont{and} \bibinfo{author}{\bibfnamefont{V.~I.} \bibnamefont{Klimov}}, \bibinfo{journal}{Accounts of chemical research} \textbf{\bibinfo{volume}{41}}, \bibinfo{pages}{1810} (\bibinfo{year}{2008}).

\bibitem[{\citenamefont{Midgett et~al.}(2010)\citenamefont{Midgett, Hillhouse, Hughes, Nozik, and Beard}}]{midgett2010flowing}
\bibinfo{author}{\bibfnamefont{A.~G.} \bibnamefont{Midgett}}, \bibinfo{author}{\bibfnamefont{H.~W.} \bibnamefont{Hillhouse}}, \bibinfo{author}{\bibfnamefont{B.~K.} \bibnamefont{Hughes}}, \bibinfo{author}{\bibfnamefont{A.~J.} \bibnamefont{Nozik}}, \bibnamefont{and} \bibinfo{author}{\bibfnamefont{M.~C.} \bibnamefont{Beard}}, \bibinfo{journal}{The Journal of Physical Chemistry C} \textbf{\bibinfo{volume}{114}}, \bibinfo{pages}{17486} (\bibinfo{year}{2010}).

\bibitem[{\citenamefont{Beard et~al.}(2010)\citenamefont{Beard, Midgett, Hanna, Luther, Hughes, and Nozik}}]{beard2010comparing}
\bibinfo{author}{\bibfnamefont{M.~C.} \bibnamefont{Beard}}, \bibinfo{author}{\bibfnamefont{A.~G.} \bibnamefont{Midgett}}, \bibinfo{author}{\bibfnamefont{M.~C.} \bibnamefont{Hanna}}, \bibinfo{author}{\bibfnamefont{J.~M.} \bibnamefont{Luther}}, \bibinfo{author}{\bibfnamefont{B.~K.} \bibnamefont{Hughes}}, \bibnamefont{and} \bibinfo{author}{\bibfnamefont{A.~J.} \bibnamefont{Nozik}}, \bibinfo{journal}{Nano letters} \textbf{\bibinfo{volume}{10}}, \bibinfo{pages}{3019} (\bibinfo{year}{2010}).

\bibitem[{\citenamefont{Sukhovatkin et~al.}(2009)\citenamefont{Sukhovatkin, Hinds, Brzozowski, and Sargent}}]{sukhovatkin2009colloidal}
\bibinfo{author}{\bibfnamefont{V.}~\bibnamefont{Sukhovatkin}}, \bibinfo{author}{\bibfnamefont{S.}~\bibnamefont{Hinds}}, \bibinfo{author}{\bibfnamefont{L.}~\bibnamefont{Brzozowski}}, \bibnamefont{and} \bibinfo{author}{\bibfnamefont{E.~H.} \bibnamefont{Sargent}}, \bibinfo{journal}{Science} \textbf{\bibinfo{volume}{324}}, \bibinfo{pages}{1542} (\bibinfo{year}{2009}).

\bibitem[{\citenamefont{Semonin et~al.}(2011)\citenamefont{Semonin, Luther, Choi, Chen, Gao, Nozik, and Beard}}]{semonin2011peak}
\bibinfo{author}{\bibfnamefont{O.~E.} \bibnamefont{Semonin}}, \bibinfo{author}{\bibfnamefont{J.~M.} \bibnamefont{Luther}}, \bibinfo{author}{\bibfnamefont{S.}~\bibnamefont{Choi}}, \bibinfo{author}{\bibfnamefont{H.-Y.} \bibnamefont{Chen}}, \bibinfo{author}{\bibfnamefont{J.}~\bibnamefont{Gao}}, \bibinfo{author}{\bibfnamefont{A.~J.} \bibnamefont{Nozik}}, \bibnamefont{and} \bibinfo{author}{\bibfnamefont{M.~C.} \bibnamefont{Beard}}, \bibinfo{journal}{Science} \textbf{\bibinfo{volume}{334}}, \bibinfo{pages}{1530} (\bibinfo{year}{2011}).

\bibitem[{\citenamefont{Kim et~al.}(2021)\citenamefont{Kim, Tran, Kim, Yoo, Oh, Kim, and Lee}}]{kim2021escalated}
\bibinfo{author}{\bibfnamefont{J.~S.} \bibnamefont{Kim}}, \bibinfo{author}{\bibfnamefont{M.~D.} \bibnamefont{Tran}}, \bibinfo{author}{\bibfnamefont{S.~T.} \bibnamefont{Kim}}, \bibinfo{author}{\bibfnamefont{D.}~\bibnamefont{Yoo}}, \bibinfo{author}{\bibfnamefont{S.-H.} \bibnamefont{Oh}}, \bibinfo{author}{\bibfnamefont{J.-H.} \bibnamefont{Kim}}, \bibnamefont{and} \bibinfo{author}{\bibfnamefont{Y.~H.} \bibnamefont{Lee}}, \bibinfo{journal}{Nano letters} \textbf{\bibinfo{volume}{21}}, \bibinfo{pages}{1976} (\bibinfo{year}{2021}).

\bibitem[{\citenamefont{Petocchi et~al.}(2019)\citenamefont{Petocchi, Beck, Ederer, and Werner}}]{petocchi2019hund}
\bibinfo{author}{\bibfnamefont{F.}~\bibnamefont{Petocchi}}, \bibinfo{author}{\bibfnamefont{S.}~\bibnamefont{Beck}}, \bibinfo{author}{\bibfnamefont{C.}~\bibnamefont{Ederer}}, \bibnamefont{and} \bibinfo{author}{\bibfnamefont{P.}~\bibnamefont{Werner}}, \bibinfo{journal}{Physical Review B} \textbf{\bibinfo{volume}{100}}, \bibinfo{pages}{075147} (\bibinfo{year}{2019}).

\bibitem[{\citenamefont{Khan et~al.}(2014)\citenamefont{Khan, Jayabalan, Chari, Pal, Porwal, Sharma, and Oak}}]{khan2014quantum}
\bibinfo{author}{\bibfnamefont{S.}~\bibnamefont{Khan}}, \bibinfo{author}{\bibfnamefont{J.}~\bibnamefont{Jayabalan}}, \bibinfo{author}{\bibfnamefont{R.}~\bibnamefont{Chari}}, \bibinfo{author}{\bibfnamefont{S.}~\bibnamefont{Pal}}, \bibinfo{author}{\bibfnamefont{S.}~\bibnamefont{Porwal}}, \bibinfo{author}{\bibfnamefont{T.~K.} \bibnamefont{Sharma}}, \bibnamefont{and} \bibinfo{author}{\bibfnamefont{S.}~\bibnamefont{Oak}}, \bibinfo{journal}{Applied Physics Letters} \textbf{\bibinfo{volume}{105}}, \bibinfo{pages}{073106} (\bibinfo{year}{2014}).

\bibitem[{\citenamefont{Khan et~al.}(2015)\citenamefont{Khan, Jayabalan, Singh, Yogi, and Chari}}]{khan2015probing}
\bibinfo{author}{\bibfnamefont{S.}~\bibnamefont{Khan}}, \bibinfo{author}{\bibfnamefont{J.}~\bibnamefont{Jayabalan}}, \bibinfo{author}{\bibfnamefont{A.}~\bibnamefont{Singh}}, \bibinfo{author}{\bibfnamefont{R.}~\bibnamefont{Yogi}}, \bibnamefont{and} \bibinfo{author}{\bibfnamefont{R.}~\bibnamefont{Chari}}, \bibinfo{journal}{Applied Physics Letters} \textbf{\bibinfo{volume}{107}}, \bibinfo{pages}{121905} (\bibinfo{year}{2015}).

\bibitem[{\citenamefont{Qazilbash et~al.}(2008)\citenamefont{Qazilbash, Schafgans, Burch, Yun, Chae, Kim, Kim, and Basov}}]{qazilbash2008electrodynamics}
\bibinfo{author}{\bibfnamefont{M.~M.} \bibnamefont{Qazilbash}}, \bibinfo{author}{\bibfnamefont{A.}~\bibnamefont{Schafgans}}, \bibinfo{author}{\bibfnamefont{K.}~\bibnamefont{Burch}}, \bibinfo{author}{\bibfnamefont{S.}~\bibnamefont{Yun}}, \bibinfo{author}{\bibfnamefont{B.}~\bibnamefont{Chae}}, \bibinfo{author}{\bibfnamefont{B.}~\bibnamefont{Kim}}, \bibinfo{author}{\bibfnamefont{H.-T.} \bibnamefont{Kim}}, \bibnamefont{and} \bibinfo{author}{\bibfnamefont{D.}~\bibnamefont{Basov}}, \bibinfo{journal}{Physical Review B} \textbf{\bibinfo{volume}{77}}, \bibinfo{pages}{115121} (\bibinfo{year}{2008}).

\bibitem[{\citenamefont{Supplementary}(2023)}]{supplementary}
\bibinfo{author}{\bibnamefont{Supplementary}}, \bibinfo{journal}{see supplementary material for raw data of spectroscopic ellipsometry, quantum efficiency calculation details, pump and probe beam spectra for ultrafast transient reflectivity experiment, and table of fitted parameters from transient reflectivity}  (\bibinfo{year}{2023}).

\bibitem[{\citenamefont{Lee et~al.}(2016)\citenamefont{Lee, Ivanov, Keum, and Lee}}]{lee2016epitaxial}
\bibinfo{author}{\bibfnamefont{S.}~\bibnamefont{Lee}}, \bibinfo{author}{\bibfnamefont{I.~N.} \bibnamefont{Ivanov}}, \bibinfo{author}{\bibfnamefont{J.~K.} \bibnamefont{Keum}}, \bibnamefont{and} \bibinfo{author}{\bibfnamefont{H.~N.} \bibnamefont{Lee}}, \bibinfo{journal}{Scientific reports} \textbf{\bibinfo{volume}{6}}, \bibinfo{pages}{1} (\bibinfo{year}{2016}).

\bibitem[{\citenamefont{Shibuya and Sawa}(2017)}]{shibuya2017polarized}
\bibinfo{author}{\bibfnamefont{K.}~\bibnamefont{Shibuya}} \bibnamefont{and} \bibinfo{author}{\bibfnamefont{A.}~\bibnamefont{Sawa}}, \bibinfo{journal}{Journal of Applied Physics} \textbf{\bibinfo{volume}{122}}, \bibinfo{pages}{015307} (\bibinfo{year}{2017}).

\bibitem[{\citenamefont{Marini et~al.}(2008)\citenamefont{Marini, Arcangeletti, Di~Castro, Baldassare, Perucchi, Lupi, Malavasi, Boeri, Pomjakushina, Conder et~al.}}]{marini2008optical}
\bibinfo{author}{\bibfnamefont{C.}~\bibnamefont{Marini}}, \bibinfo{author}{\bibfnamefont{E.}~\bibnamefont{Arcangeletti}}, \bibinfo{author}{\bibfnamefont{D.}~\bibnamefont{Di~Castro}}, \bibinfo{author}{\bibfnamefont{L.}~\bibnamefont{Baldassare}}, \bibinfo{author}{\bibfnamefont{A.}~\bibnamefont{Perucchi}}, \bibinfo{author}{\bibfnamefont{S.}~\bibnamefont{Lupi}}, \bibinfo{author}{\bibfnamefont{L.}~\bibnamefont{Malavasi}}, \bibinfo{author}{\bibfnamefont{L.}~\bibnamefont{Boeri}}, \bibinfo{author}{\bibfnamefont{E.}~\bibnamefont{Pomjakushina}}, \bibinfo{author}{\bibfnamefont{K.}~\bibnamefont{Conder}}, \bibnamefont{et~al.}, \bibinfo{journal}{Physical Review B} \textbf{\bibinfo{volume}{77}}, \bibinfo{pages}{235111} (\bibinfo{year}{2008}).

\bibitem[{\citenamefont{Chen}(2011)}]{chen2011assignment}
\bibinfo{author}{\bibfnamefont{X.-B.} \bibnamefont{Chen}}, \bibinfo{journal}{Journal of the Korean Physical Society} \textbf{\bibinfo{volume}{58}}, \bibinfo{pages}{100} (\bibinfo{year}{2011}).

\bibitem[{\citenamefont{Kana et~al.}(2011)\citenamefont{Kana, Ndjaka, Vignaud, Gibaud, and Maaza}}]{kana2011thermally}
\bibinfo{author}{\bibfnamefont{J.~K.} \bibnamefont{Kana}}, \bibinfo{author}{\bibfnamefont{J.}~\bibnamefont{Ndjaka}}, \bibinfo{author}{\bibfnamefont{G.}~\bibnamefont{Vignaud}}, \bibinfo{author}{\bibfnamefont{A.}~\bibnamefont{Gibaud}}, \bibnamefont{and} \bibinfo{author}{\bibfnamefont{M.}~\bibnamefont{Maaza}}, \bibinfo{journal}{Optics Communications} \textbf{\bibinfo{volume}{284}}, \bibinfo{pages}{807} (\bibinfo{year}{2011}).

\bibitem[{\citenamefont{Huffman et~al.}(2017)\citenamefont{Huffman, Hendriks, Walter, Yoon, Ju, Smith, Carr, Krakauer, and Qazilbash}}]{huffman2017insulating}
\bibinfo{author}{\bibfnamefont{T.}~\bibnamefont{Huffman}}, \bibinfo{author}{\bibfnamefont{C.}~\bibnamefont{Hendriks}}, \bibinfo{author}{\bibfnamefont{E.}~\bibnamefont{Walter}}, \bibinfo{author}{\bibfnamefont{J.}~\bibnamefont{Yoon}}, \bibinfo{author}{\bibfnamefont{H.}~\bibnamefont{Ju}}, \bibinfo{author}{\bibfnamefont{R.}~\bibnamefont{Smith}}, \bibinfo{author}{\bibfnamefont{G.}~\bibnamefont{Carr}}, \bibinfo{author}{\bibfnamefont{H.}~\bibnamefont{Krakauer}}, \bibnamefont{and} \bibinfo{author}{\bibfnamefont{M.}~\bibnamefont{Qazilbash}}, \bibinfo{journal}{Physical Review B} \textbf{\bibinfo{volume}{95}}, \bibinfo{pages}{075125} (\bibinfo{year}{2017}).

\bibitem[{\citenamefont{Qi et~al.}(2005)\citenamefont{Qi, Fischbein, Drndi{\'c}, and {\v{S}}elmi{\'c}}}]{qi2005efficient}
\bibinfo{author}{\bibfnamefont{D.}~\bibnamefont{Qi}}, \bibinfo{author}{\bibfnamefont{M.}~\bibnamefont{Fischbein}}, \bibinfo{author}{\bibfnamefont{M.}~\bibnamefont{Drndi{\'c}}}, \bibnamefont{and} \bibinfo{author}{\bibfnamefont{S.}~\bibnamefont{{\v{S}}elmi{\'c}}}, \bibinfo{journal}{Applied Physics Letters} \textbf{\bibinfo{volume}{86}}, \bibinfo{pages}{093103} (\bibinfo{year}{2005}).

\bibitem[{\citenamefont{Gao et~al.}(2015)\citenamefont{Gao, Fidler, and Klimov}}]{gao2015carrier}
\bibinfo{author}{\bibfnamefont{J.}~\bibnamefont{Gao}}, \bibinfo{author}{\bibfnamefont{A.~F.} \bibnamefont{Fidler}}, \bibnamefont{and} \bibinfo{author}{\bibfnamefont{V.~I.} \bibnamefont{Klimov}}, \bibinfo{journal}{Nature Communications} \textbf{\bibinfo{volume}{6}}, \bibinfo{pages}{8185} (\bibinfo{year}{2015}).

\bibitem[{\citenamefont{Villamil~Franco et~al.}(2020)\citenamefont{Villamil~Franco, Mahler, Cornaggia, Gustavsson, and Cassette}}]{villamil2020auger}
\bibinfo{author}{\bibfnamefont{C.}~\bibnamefont{Villamil~Franco}}, \bibinfo{author}{\bibfnamefont{B.}~\bibnamefont{Mahler}}, \bibinfo{author}{\bibfnamefont{C.}~\bibnamefont{Cornaggia}}, \bibinfo{author}{\bibfnamefont{T.}~\bibnamefont{Gustavsson}}, \bibnamefont{and} \bibinfo{author}{\bibfnamefont{E.}~\bibnamefont{Cassette}}, \bibinfo{journal}{ACS Applied Nano Materials} \textbf{\bibinfo{volume}{4}}, \bibinfo{pages}{558} (\bibinfo{year}{2020}).

\bibitem[{\citenamefont{Caruthers and Kleinman}(1973)}]{caruthers1973energy}
\bibinfo{author}{\bibfnamefont{E.}~\bibnamefont{Caruthers}} \bibnamefont{and} \bibinfo{author}{\bibfnamefont{L.}~\bibnamefont{Kleinman}}, \bibinfo{journal}{Physical Review B} \textbf{\bibinfo{volume}{7}}, \bibinfo{pages}{3760} (\bibinfo{year}{1973}).

\bibitem[{\citenamefont{Gavini and Kwan}(1972)}]{gavini1972optical}
\bibinfo{author}{\bibfnamefont{A.}~\bibnamefont{Gavini}} \bibnamefont{and} \bibinfo{author}{\bibfnamefont{C.~C.} \bibnamefont{Kwan}}, \bibinfo{journal}{Physical Review B} \textbf{\bibinfo{volume}{5}}, \bibinfo{pages}{3138} (\bibinfo{year}{1972}).

\bibitem[{\citenamefont{Verleur et~al.}(1968)\citenamefont{Verleur, Barker~Jr, and Berglund}}]{verleur1968optical}
\bibinfo{author}{\bibfnamefont{H.~W.} \bibnamefont{Verleur}}, \bibinfo{author}{\bibfnamefont{A.}~\bibnamefont{Barker~Jr}}, \bibnamefont{and} \bibinfo{author}{\bibfnamefont{C.}~\bibnamefont{Berglund}}, \bibinfo{journal}{Physical Review} \textbf{\bibinfo{volume}{172}}, \bibinfo{pages}{788} (\bibinfo{year}{1968}).

\bibitem[{\citenamefont{Dai et~al.}(2019)\citenamefont{Dai, Lian, Miller, Wang, Shi, Liu, Song, and Wang}}]{dai2019optical}
\bibinfo{author}{\bibfnamefont{K.}~\bibnamefont{Dai}}, \bibinfo{author}{\bibfnamefont{J.}~\bibnamefont{Lian}}, \bibinfo{author}{\bibfnamefont{M.~J.} \bibnamefont{Miller}}, \bibinfo{author}{\bibfnamefont{J.}~\bibnamefont{Wang}}, \bibinfo{author}{\bibfnamefont{Y.}~\bibnamefont{Shi}}, \bibinfo{author}{\bibfnamefont{Y.}~\bibnamefont{Liu}}, \bibinfo{author}{\bibfnamefont{H.}~\bibnamefont{Song}}, \bibnamefont{and} \bibinfo{author}{\bibfnamefont{X.}~\bibnamefont{Wang}}, \bibinfo{journal}{Optical Materials Express} \textbf{\bibinfo{volume}{9}}, \bibinfo{pages}{663} (\bibinfo{year}{2019}).

\bibitem[{\citenamefont{Schneider}(2020)}]{schneider2020optical}
\bibinfo{author}{\bibfnamefont{K.}~\bibnamefont{Schneider}}, \bibinfo{journal}{Journal of Materials Science: Materials in Electronics} \textbf{\bibinfo{volume}{31}}, \bibinfo{pages}{10478} (\bibinfo{year}{2020}).

\bibitem[{\citenamefont{Pelant and Valenta}(2012)}]{pelant2012luminescence}
\bibinfo{author}{\bibfnamefont{I.}~\bibnamefont{Pelant}} \bibnamefont{and} \bibinfo{author}{\bibfnamefont{J.}~\bibnamefont{Valenta}}, \emph{\bibinfo{title}{Luminescence spectroscopy of semiconductors}} (\bibinfo{publisher}{OUP Oxford}, \bibinfo{year}{2012}).

\bibitem[{\citenamefont{Holleman et~al.}(2016)\citenamefont{Holleman, Bishop, Garcia, Winfred, Lee, Lee, Beekman, Manousakis, and McGill}}]{holleman2016evidence}
\bibinfo{author}{\bibfnamefont{J.}~\bibnamefont{Holleman}}, \bibinfo{author}{\bibfnamefont{M.~M.} \bibnamefont{Bishop}}, \bibinfo{author}{\bibfnamefont{C.}~\bibnamefont{Garcia}}, \bibinfo{author}{\bibfnamefont{J.~V.} \bibnamefont{Winfred}}, \bibinfo{author}{\bibfnamefont{S.}~\bibnamefont{Lee}}, \bibinfo{author}{\bibfnamefont{H.~N.} \bibnamefont{Lee}}, \bibinfo{author}{\bibfnamefont{C.}~\bibnamefont{Beekman}}, \bibinfo{author}{\bibfnamefont{E.}~\bibnamefont{Manousakis}}, \bibnamefont{and} \bibinfo{author}{\bibfnamefont{S.~A.} \bibnamefont{McGill}}, \bibinfo{journal}{Physical Review B} \textbf{\bibinfo{volume}{94}}, \bibinfo{pages}{155129} (\bibinfo{year}{2016}).

\bibitem[{\citenamefont{Okazaki et~al.}(2002)\citenamefont{Okazaki, Fujimori, and Onoda}}]{okazaki2002temperature}
\bibinfo{author}{\bibfnamefont{K.}~\bibnamefont{Okazaki}}, \bibinfo{author}{\bibfnamefont{A.}~\bibnamefont{Fujimori}}, \bibnamefont{and} \bibinfo{author}{\bibfnamefont{M.}~\bibnamefont{Onoda}}, \bibinfo{journal}{Journal of the Physical Society of Japan} \textbf{\bibinfo{volume}{71}}, \bibinfo{pages}{822} (\bibinfo{year}{2002}).

\bibitem[{\citenamefont{Wang and Gao}(2015)}]{wang2015distinguishing}
\bibinfo{author}{\bibfnamefont{X.}~\bibnamefont{Wang}} \bibnamefont{and} \bibinfo{author}{\bibfnamefont{H.}~\bibnamefont{Gao}}, \bibinfo{journal}{Nano letters} \textbf{\bibinfo{volume}{15}}, \bibinfo{pages}{7037} (\bibinfo{year}{2015}).

\end{thebibliography}
\end{document}